\documentclass[aip,reprint]{revtex4-1} 

\usepackage{natbib}
\usepackage[version=3]{mhchem} 
\usepackage{bm} 
\usepackage{threeparttable} 
\usepackage{dcolumn}
	\newcolumntype{d}[1]{D{.}{.}{#1}}
\usepackage{booktabs} 
\usepackage{xspace}
\usepackage{braket}
\usepackage{comment}
\usepackage{color, colortbl}
\usepackage{soul}
\usepackage{titlesec}
\titlelabel{\thetitle.\quad}
\usepackage{extarrows}
\usepackage{mathtools}

\usepackage{todonotes}

\usepackage{hyperref} 

\newcommand*{\etal}{\xspace et al.\xspace}
\DeclareMathOperator{\ext}{ext}
\newcommand{\gbk}[2]{\ensuremath{\left( #1 \middle| #2 \right)}}
\newcommand{\thws}{\ensuremath{\vartheta_{\mathrm{ws}}}\xspace}
\newcommand{\thsq}{\ensuremath{\vartheta_{\mathrm{SQ}}}\xspace}
\newcommand{\sqvl}{SQV$\ell$\xspace}

\begin{document}


\title{Comment on ``A tight distance-dependent estimator for screening three-center Coulomb integrals over Gaussian basis functions''  [J. Chem. Phys. 142, 154106 (2015)]} 

\author{Edward~F.~Valeev}
\email{efv@vt.edu}
\affiliation{Department of Chemistry, Virginia Tech, Blacksburg, Virginia 24061, USA}

\author{Toru Shiozaki}
\affiliation{Quantum Simulation Technologies, Inc., Cambridge, MA 02139, USA}

\date{\today}


\maketitle 

Hollman\etal presented a set of formulas\cite{Hollman:2015ca} for estimating the magnitude of 3-center Coulomb integrals over (contracted, solid-harmonic) Gaussian atomic orbitals (AO). Evaluation of such integrals 
often accounts for a significant percentage of the computational cost of electronic structure methods that utilize density fitting ({\em aka} the resolution-of-the-identity) of the Coulomb Hamiltonian.
Unlike the standard Cauchy-Schwarz bounds for the AO integrals, the estimators of Hollman\etal do not produce
upper bounds but correctly describe the asymptotic decay of the integral with the distance between the (chemist's) bra and ket.
Namely, for well-separated bra $\phi_a ({\bf r}_1) \phi_b({\bf r}_1)$ and ket $\phi_c({\bf r}_2)$ the corresponding
Coulomb integral $(\phi_a \phi_b| \phi_c)$ (the notation of Ref. \onlinecite{Hollman:2015ca} is used throughout)
generally decays with the distance $R$ between the centers-of-charge of bra and ket as $R^{l_c+1}$.
The estimators of Hollman\etal incorporate this distance dependence correctly, whereas the Cauchy-Schwarz
counterpart does not depend on $R$ thus severely overestimating the magnitude of the vast majority of
nonnegligible 3-center integrals.

An even faster decay is observed for three-center Coulomb integrals in which the AOs in the $(\phi_a \phi_b|$ bra are (a) {\em concentric}, i.e., share the origin, and have different angular momenta, or (b) one of the two functions
is the identity. Unfortunately, Hollman\etal did not provide formulas for such {\em two-center} integrals, stating simply:
``A better estimate for the special two-center case could be developed, but from a practical standpoint it is not worth the effort.''
Although the number and cost of such two-center Coulomb integrals is indeed smaller than that of three-center integrals, their cost is non-negligible.
Since the utility of the two-center integrals extends beyond the density fitting to semiempirical and classical simulations their efficient evaluation
has recently been revisited\cite{Peels:2020gh}.
A priori estimator for the magnitudes of such integrals that incorporates the proper decay with distance would therefore be useful.
Albeit the three-center estimators cannot be trivially adapted to the two-center case, the derivation is straightforward and does not warrant a full manuscript.
The purpose of this comment is to fill this gap in Ref. \onlinecite{Hollman:2015ca} by documenting the appropriate two-center formulas. We also propose a modification
of the three-center estimator of Hollman\etal for the case of the contracted ket AO.

Consider a two-center Coulomb integral, $(\phi_a| \phi_b)$ involving two solid-harmonic Gaussian AOs $\phi_a$ and $\phi_b$ with orbital angular quanta $l_a$ and $l_b$.
When $\phi_a$ and $\phi_b$ are well-separated (see Ref. \onlinecite{Hollman:2015ca} for the definitions of orbital extents and well-separatedness; the use of recently-proposed improvements for the extent definitions by Ochsenfeld and co-workers\cite{Thompson:2019ch} could lead to further improvements)
the integral decays with the distance between their origins, $R$, as $R^{l_a+l_b+1}$.
It is sufficient to consider the case with the AO origins on the z axis and solid-harmonic Gaussians $\phi_a$ and $\phi_b$ with null orbital angular momentum
projection ($m=0$); this makes all spherical electric multipole moments $O^l_m$ of AOs $\phi_a$ and $\phi_b$ vanish except 
$O^{l_a}_0 (\phi_a)$ and $O^{l_b}_0 (\phi_b)$, respectively.
Then the bipolar multipole expansion\cite{Carlson:1950ca,Rose:1958cn} of  $(\phi_a| \phi_b)$ reduces to the single term:
\begin{align}
(\phi_a | \phi_b) = & (-1)^{l_b} \sqrt{ \begin{pmatrix} l_a + l_b \\ l_a \end{pmatrix} }  \begin{pmatrix} l_a & l_b & l_a+l_b \\ 0 & 0 & 0 \end{pmatrix}  \frac{O^{l_a}_0(\phi_a) O^{l_b}_0(\phi_b)}{R^{l_a + l_b + 1}}
\end{align}
After straightforward simplification this formula can be immediately used to efficiently estimate $|(\phi_a | \phi_b)|$ for well-separated
contracted Gaussian AOs; for integrals with overlapping brakets Cauchy-Schwarz estimator should be used, just as in Ref. \onlinecite{Hollman:2015ca}. This results in the proposed
two-center estimator:
\begin{align}
\label{eq:ab}
|(\phi_a | \phi_b)| \overset{\text{SQV}\ell}{\approx} &
\begin{dcases}
    \begin{pmatrix} l_a + l_b \\ l_a \end{pmatrix} \frac{O^{l_a}_0(\phi_a) O^{l_b}_0(\phi_b)}{R^{l_a + l_b + 1}}, & \!\!R > \ext_{a} + \ext_{b}, \\
    Q_{a}Q_b, & \!\!R \leq \ext_{a} + \ext_{b},
  \end{dcases}
\end{align}
where the Cauchy-Schwarz parameters $Q$ and extents were introduced as in Ref. \onlinecite{Hollman:2015ca} (note that the definition of the former
in numerical examples shown below differ from that of Hollman et al. as discussed below).
For efficient use the combinatorial prefactor
and multipole moments of Gaussian AOs should be pretabulated (multipole moment integrals can be evaluated, e.g., recursively\cite{PerezJorda:1996jj} (mind the several conventions for the multipole moments in the literature) or using the contracted version of Eq. \eqref{eq:O}).

Although the well-separated clause Eq. \eqref{eq:ab} does not seem to resemble its counterparts in the
three-center estimator of Ref. \onlinecite{Hollman:2015ca} [(Eq. (29)], their connection becomes apparent if we recall that
the standard-convention multipole moment $O^{l}_0(\phi) \equiv \int d\,\mathbf{r} \, \phi(\mathbf{r}) r^l P_l(\cos(\theta))$  of a {\em primitive} solid-harmonic Gaussian AO $\phi$ with orbital exponent $\zeta$ and angular momentum $l$ has a compact expression
in terms of the $\beta_l(\zeta) \equiv (2l-1)!! \zeta^{-(2l+3)/4}$ function introduced in Eq. (28) of Ref. \onlinecite{Hollman:2015ca}:
\begin{align}
\label{eq:O}
O^{l}_0(\phi) = (2 \pi)^{3/4} \beta_l(\zeta)
\end{align}
With this equality we can clearly identify the first (SV$\ell$) clause of Eq. (29) in Ref. \onlinecite{Hollman:2015ca} as simply the interaction of the {\em charge} of the $(ab|$ bra
with the lone surviving multipole of $|c)$ ket. This perspective also allows to make the three-center estimator
more sound for the case of {\em contracted} $|c)$, in which case the prescription of Hollman\etal in Section II.F
approximates the sole nonvanishing multipole moment $O^{l_c}_0(\phi_c)$ by its value for the most-diffuse primitive only.
It is more sound to use the {\em exact} multipole moment for the contracted $\phi_c$, achieved
by replacing  $\beta_{l_c}(\zeta_c)$ with $O^{l_c}_0(\phi_c) (2 \pi)^{-3/4}$ in Eq. (29) of Ref. \onlinecite{Hollman:2015ca}.
If the density fitting AOs are not contracted the amendment has no effect, however for fitting sets employing
contracted Gaussians the amended form of \sqvl  should be preferred.
To gauge the performance of \sqvl employing the exact multipole moment for contracted $\phi_c$
we reassessed the performance of \sqvl for the {\em well-separated} 3-center Coulomb integrals
in the same benchmark set of molecules
used in Ref. \onlinecite{Hollman:2015ca}
with def2-SVP/def2-SVP/C orbital/fitting basis set pair.
Since the def2-SVP/C fitting basis includes contracted basis functions the original and amended \sqvl
indeed produce different results, as shown in Table \ref{tab:table1} (see Ref. \onlinecite{Hollman:2015ca}
for the definitions of statistical measures). The use of exact multipole moment for contracted $\phi_c$
brings the integral estimates closer to the actual values.

Note that the statistics reported for
the original \sqvl differs from that given in Table I in Ref. \onlinecite{Hollman:2015ca}
because a different definition of
the Cauchy-Schwarz parameters $Q$ was utilized in that work; namely, $Q_{ab}$ for a shell-pair
was defined by Hollman et al. (Eq. (24) in Ref. \onlinecite{Hollman:2015ca})
as the Frobenius norm of the ``diagonal'' of the $(\phi_a \phi_b|\phi_a \phi_b)$ shell
quartet. This was done to make the Schwarz estimate lab frame invariant (unlike the traditional infinity-norm-based
use of the Schwarz inequality due to its dependence on the orientation of the individual AOs).
But such measure of significance of $\phi_a \phi_b$ is clearly
not a norm, and cannot be used in Cauchy-Schwarz inequality.
Here we define $Q_{ab}$ and $Q_c$ as the Frobenius norm
of shell quartet $(\phi_a \phi_b|\phi_a \phi_b)$ and shell doublet
$(\phi_c|\phi_c)$, respectively; such choice makes Schwarz and \sqvl estimates
lab frame invariant and ensures the upper-bound property of the Schwarz estimates.

\begin{widetext}
\begin{center}
\begin{table}[htp]
\caption{Statistical measures of the original and improved variants of the \sqvl estimator. See Ref. \onlinecite{Hollman:2015ca}
for the definitions of statistical measures. The def2-SVP/def2-SVP/C orbital/fitting basis set pair was used. The \sqvl parameters $\thws$ and $\thsq$ were set to 0.1.} \label{tab:table1}
\begin{tabular}{c|c|cccccc}
\hline\hline
concentric $\phi_a \phi_b$ excluded? & \sqvl variant & $\bar{F}$ & $\sigma(\log F)$ & $F_\text{max}$ & $F_\text{min}$ & $N_\text{ws}/10^6$ \\ \hline
yes & original$^*$ &
1.434 & 0.483 & 20.615 & 0.103 & 26.2 \\
yes & amended & 0.984 & 0.123 & 3.860 & 0.067 & 26.2 \\ \hline
no & original &
2.394 & 0.617 & 6545 & 0.103 & 27.6 \\ 
no & amended & 0.977 & 0.139 & 3.860 & 0.067 & 27.6
\\
\hline\hline
\end{tabular}
\\
$^*$ The data differs from that in Table I of Ref. \onlinecite{Hollman:2015ca}; see the text for details.
\label{default}
\end{table}%
\end{center}
\end{widetext}

Lastly, the 2-center \sqvl (Eq. \eqref{eq:ab}) offers yet another way to improve the \sqvl estimator for 3-center Coulomb integrals,
 by allowing to estimate $(\phi_a \phi_b|\phi_c)$ more accurately when $\phi_a$ and $\phi_b$ are concentric .
 In such case the product density $\phi_a \phi_b$ has a leading nonvanishing moment of order $|l_a - l_b|$, i.e., nonzero if the angular momenta of $\phi_a$ and $\phi_b$
 differ (product of two Cartesian Gaussian AOs has zero/nonzero charge if the total Cartesian quanta of the two functions have different/same parities).
 The original \sqvl estimator severely overestimates such integrals, albeit this fact was obscured in Ref. \onlinecite{Hollman:2015ca}
 by excluding integrals with concentric $\phi_a \phi_b$ from the statistics.
 To account for the faster decay for many well-separated brakets we can use  Eq. \eqref{eq:ab} directly to handle the concentric bra case.
 Of course, this is only done for integrals with well-separated brakets.
 
Combining the proposed two improvements we obtain the following {\em amended} \sqvl estimator:
\begin{widetext}
\begin{align}
  \left|\gbk{\phi_a \phi_b}{\phi_c}\right| \overset{\text{SQV}\ell}{\approx} 
  \begin{dcases}
    \begin{pmatrix} |l_a - l_b| + l_c \\ l_c \end{pmatrix} \frac{O^{|l_a-l_b|}_0(\phi_a \phi_b) O^{l_c}_0(\phi_c)}{R^{|l_a - l_b| + l_c + 1}}, & \!\!\begin{multlined}[c]
      R > \ext_{ab} + \ext_{c}  \\ \text{and } \phi_a \text{ and } \phi_b \, \text{are\ concentric}
    \end{multlined} 
    \\
    |S_{ab}|\frac{O^{l_c}_0(\phi_c)}{R^{l_c+1}}
    & \!\!\begin{multlined}[c]
      R > \ext_{ab} + \ext_{c}  \\ \text{and } S_{ab}/Q_{ab} > \thsq 
    \end{multlined} 
    \\
    \frac{Q_{ab} \, \pi^{1/4}}{\left(2(\zeta_a+\zeta_b)\right)^{1/4}} \frac{O^{l}_0(\phi_c)}{R^{l_c+1}}
    & \!\!\begin{multlined}[c]
      R > \ext_{ab} + \ext_{c} \\ \text{and } S_{ab}/Q_{ab} \leq \thsq 
    \end{multlined} 
    \\
    \ Q_{ab}Q_c & \!\! R \leq \ext_{ab} + \ext_{c}.
  \end{dcases}
  \label{eq:sqvl}
\end{align}
\end{widetext}
Note that the estimate for the concentric case involves $O^{|l_a-l_b|}_0(\phi_a \phi_b)$, the spherical electric multipole moment of {\em product} density $\phi_a \phi_b$
that can be evaluated recursively\cite{PerezJorda:1996jj} or, for the case of primitive solid-harmonic AOs $\phi_a$ and $\phi_b$ with null orbital angular momentum
projection ($m=0$), explicit formula can be used:
\begin{widetext}
\begin{align}
O^{|l_a-l_b|}_0(\phi_a \phi_b) = & \binom{l_a}{l_b}
    (2 |l_a- l_b|-1)!!   
         \sqrt{\frac{(2 l_b-1)!!}{(2
   l_a-1)!!}}
   \left( 2 \sqrt{\zeta
   _b}\right)^{l_b + \frac{3}{2}}
   \left(\frac{\sqrt{\zeta_a}}{\zeta_a
   +\zeta_b}\right)^{l_a+\frac{3}{2}},
   \end{align}
\end{widetext}
where $l_a \geq l_b$ was assumed; extensions to contracted AOs and Cartesian AOs (if needed) are trivial.
As shown in Table I the performance of the \sqvl equipped with the concentric clause
indeed is significantly improved: the severe overestimation of some well-separated integrals with concentric bras
by the original \sqvl estimator (obscured in
Ref. \onlinecite{Hollman:2015ca} by excluding such integrals when computing statistics)
is rectified by the amended \sqvl estimator.

In summary, this Comment documents a new 2-center extension of the \sqvl estimator of Hollman et al.\cite{Hollman:2015ca}
as well as nearly-cost-free improvements the original 3-center \sqvl estimator for the $\gbk{\phi_a \phi_b}{\phi_c}$ integrals
with contracted ket $\phi_c$
and/or concentric bra $\phi_a \phi_b$.


This work was supported by the U.S. National Science Foundation (awards 1550456 and 1800348).

%

\bibliography{refs}

\begin{thebibliography}{6}%
\makeatletter
\providecommand \@ifxundefined [1]{%
 \@ifx{#1\undefined}
}%
\providecommand \@ifnum [1]{%
 \ifnum #1\expandafter \@firstoftwo
 \else \expandafter \@secondoftwo
 \fi
}%
\providecommand \@ifx [1]{%
 \ifx #1\expandafter \@firstoftwo
 \else \expandafter \@secondoftwo
 \fi
}%
\providecommand \natexlab [1]{#1}%
\providecommand \enquote  [1]{``#1''}%
\providecommand \bibnamefont  [1]{#1}%
\providecommand \bibfnamefont [1]{#1}%
\providecommand \citenamefont [1]{#1}%
\providecommand \href@noop [0]{\@secondoftwo}%
\providecommand \href [0]{\begingroup \@sanitize@url \@href}%
\providecommand \@href[1]{\@@startlink{#1}\@@href}%
\providecommand \@@href[1]{\endgroup#1\@@endlink}%
\providecommand \@sanitize@url [0]{\catcode `\\12\catcode `\$12\catcode
  `\&12\catcode `\#12\catcode `\^12\catcode `\_12\catcode `\%12\relax}%
\providecommand \@@startlink[1]{}%
\providecommand \@@endlink[0]{}%
\providecommand \url  [0]{\begingroup\@sanitize@url \@url }%
\providecommand \@url [1]{\endgroup\@href {#1}{\urlprefix }}%
\providecommand \urlprefix  [0]{URL }%
\providecommand \Eprint [0]{\href }%
\providecommand \doibase [0]{http://dx.doi.org/}%
\providecommand \selectlanguage [0]{\@gobble}%
\providecommand \bibinfo  [0]{\@secondoftwo}%
\providecommand \bibfield  [0]{\@secondoftwo}%
\providecommand \translation [1]{[#1]}%
\providecommand \BibitemOpen [0]{}%
\providecommand \bibitemStop [0]{}%
\providecommand \bibitemNoStop [0]{.\EOS\space}%
\providecommand \EOS [0]{\spacefactor3000\relax}%
\providecommand \BibitemShut  [1]{\csname bibitem#1\endcsname}%
\let\auto@bib@innerbib\@empty
\bibitem [{\citenamefont {Hollman}, \citenamefont {Schaefer},\ and\
  \citenamefont {Valeev}(2015)}]{Hollman:2015ca}%
  \BibitemOpen
  \bibfield  {author} {\bibinfo {author} {\bibfnamefont {D.~S.}\ \bibnamefont
  {Hollman}}, \bibinfo {author} {\bibfnamefont {H.~F.}\ \bibnamefont
  {Schaefer}}, \ and\ \bibinfo {author} {\bibfnamefont {E.~F.}\ \bibnamefont
  {Valeev}},\ }\bibfield  {title} {\enquote {\bibinfo {title} {{A tight
  distance-dependent estimator for screening three-center Coulomb integrals
  over Gaussian basis functions}},}\ }\href@noop {} {\bibfield  {journal}
  {\bibinfo  {journal} {J Chem Phys}\ }\textbf {\bibinfo {volume} {142}},\
  \bibinfo {pages} {154106} (\bibinfo {year} {2015})}\BibitemShut {NoStop}%
\bibitem [{\citenamefont {Peels}\ and\ \citenamefont
  {Knizia}(2020)}]{Peels:2020gh}%
  \BibitemOpen
  \bibfield  {author} {\bibinfo {author} {\bibfnamefont {M.}~\bibnamefont
  {Peels}}\ and\ \bibinfo {author} {\bibfnamefont {G.}~\bibnamefont {Knizia}},\
  }\bibfield  {title} {\enquote {\bibinfo {title} {{Fast Evaluation of
  Two-Center Integrals over Gaussian Charge Distributions and Gaussian Orbitals
  with General Interaction Kernels}},}\ }\href@noop {} {\bibfield  {journal}
  {\bibinfo  {journal} {J. Chem. Theory Comput.}\ }\textbf {\bibinfo {volume}
  {16}},\ \bibinfo {pages} {2570--2583} (\bibinfo {year} {2020})}\BibitemShut
  {NoStop}%
\bibitem [{\citenamefont {Thompson}\ and\ \citenamefont
  {Ochsenfeld}(2019)}]{Thompson:2019ch}%
  \BibitemOpen
  \bibfield  {author} {\bibinfo {author} {\bibfnamefont {T.~H.}\ \bibnamefont
  {Thompson}}\ and\ \bibinfo {author} {\bibfnamefont {C.}~\bibnamefont
  {Ochsenfeld}},\ }\bibfield  {title} {\enquote {\bibinfo {title} {{Integral
  partition bounds for fast and effective screening of general one-, two-, and
  many-electron integrals}},}\ }\href@noop {} {\bibfield  {journal} {\bibinfo
  {journal} {J Chem Phys}\ }\textbf {\bibinfo {volume} {150}},\ \bibinfo
  {pages} {044101} (\bibinfo {year} {2019})}\BibitemShut {NoStop}%
\bibitem [{\citenamefont {Carlson}\ and\ \citenamefont
  {Rushbrooke}(1950)}]{Carlson:1950ca}%
  \BibitemOpen
  \bibfield  {author} {\bibinfo {author} {\bibfnamefont {B.~C.}\ \bibnamefont
  {Carlson}}\ and\ \bibinfo {author} {\bibfnamefont {G.~S.}\ \bibnamefont
  {Rushbrooke}},\ }\bibfield  {title} {\enquote {\bibinfo {title} {{On the
  expansion of a Coulomb potential in spherical harmonics}},}\ }\href@noop {}
  {\bibfield  {journal} {\bibinfo  {journal} {Math. Proc. Camb. Phil. Soc.}\
  }\textbf {\bibinfo {volume} {46}},\ \bibinfo {pages} {626--633} (\bibinfo
  {year} {1950})}\BibitemShut {NoStop}%
\bibitem [{\citenamefont {Rose}(1958)}]{Rose:1958cn}%
  \BibitemOpen
  \bibfield  {author} {\bibinfo {author} {\bibfnamefont {M.~E.}\ \bibnamefont
  {Rose}},\ }\bibfield  {title} {\enquote {\bibinfo {title} {{The Electrostatic
  Interaction of Two Arbitrary Charge Distributions}},}\ }\href@noop {}
  {\bibfield  {journal} {\bibinfo  {journal} {Journal of Mathematics and
  Physics}\ }\textbf {\bibinfo {volume} {37}},\ \bibinfo {pages} {215--222}
  (\bibinfo {year} {1958})}\BibitemShut {NoStop}%
\bibitem [{\citenamefont {P{\'e}rez-Jord{\'a}}\ and\ \citenamefont
  {Yang}(1996)}]{PerezJorda:1996jj}%
  \BibitemOpen
  \bibfield  {author} {\bibinfo {author} {\bibfnamefont {J.~M.}\ \bibnamefont
  {P{\'e}rez-Jord{\'a}}}\ and\ \bibinfo {author} {\bibfnamefont
  {W.}~\bibnamefont {Yang}},\ }\bibfield  {title} {\enquote {\bibinfo {title}
  {{A concise redefinition of the solid spherical harmonics and its use in fast
  multipole methods}},}\ }\href@noop {} {\bibfield  {journal} {\bibinfo
  {journal} {J Chem Phys}\ }\textbf {\bibinfo {volume} {104}},\ \bibinfo
  {pages} {8003--8006} (\bibinfo {year} {1996})}\BibitemShut {NoStop}%
\end{thebibliography}%

\end{document}